# Econophysics Beyond General Equilibrium: the Business Cycle Model

Victor Olkhov

TVEL, Kashirskoe sh. 49, Moscow, 115409, Russia

victor.olkhov@gmail.com

**Abstract**

Current business cycle theory is an application of the general equilibrium theory. This paper presents the business cycle model without using general equilibrium framework. We treat agents risk assessments as their coordinates *x* on economic space and establish distribution of all economic agents by their risk coordinates. We suggest aggregation of agents and their variables by scales large to compare with risk scales of single agents and small to compare with economic domain on economic space. Such model is *alike to* transition from kinetic description of multi-particle system to hydrodynamic approximation. Aggregates of agents extensive variables with risk coordinate *x* determine macro variables as functions of *x alike to* hydrodynamic variables. Economic and financial transactions between agents define evolution of their variables. Aggregation of transactions between agents with risk coordinates *x* and *y* determine macro transactions as functions of *x* and *y* and define evolution of macro variables at points *x* and *y*. We describe evolution and interactions between macro transactions by hydrodynamic-like system of economic equations. We show that business cycles are described as consequence of the system of economic equations on macro transactions. As example we describe Credit transactions *CL(t,x,y)* that provide Loans from Creditors at point *x* to Borrowers at point *y* and Loan-Repayment transactions *LR(t,x,y)* that describe repayments from Borrowers at point *y* to Creditors at point *x*. We use hydrodynamic-like economic equations and derive from them the system of ordinary differential equations that describe business cycle fluctuations of macro Credits *C(t)* and macro Loan-Repayments *LR(t)* of the entire economics. The nature of business cycle fluctuations is explained as oscillations of "mean risk" of economic variables on bounded economic domain of economic space. Our model can describe business cycle fluctuations for any number of macroeconomic and financial variables.

This research did not receive any specific grant or financial support from TVEL or funding agencies in the public, commercial, or not-for-profit sectors.

# 1. Introduction.

Macroeconomics variables follow fluctuations governed by business cycles. For decades description of business cycles remains as core macroeconomic problem [1-18]. Due to [1]: "Serious efforts to explain business crises and depressions began amid the violent fluctuations in trade which followed the Napoleonic Wars. For a century or more Western Europe had been experiencing at intervals speculative manias, glutted markets, and epidemics of bankruptcy. The Mississippi Bubble and the South Sea Scheme which had burst in France and England in 1720, and the commercial crises of 1763, 1772, 1783 and 1793, not to mention less notable cases, had excited much discussion". "The incorporation of cyclical phenomena into the system of economic equilibrium with which they are in apparent contradiction, remains the crucial problem of Trade Cycle Theory [19-20]. "Why aggregate variables undergo repeated fluctuations about trend, all of essentially the same character? Prior to Keynes' General Theory, the resolution of this question was regarded as one of the main outstanding challenges to economic research, and attempts to meet this challenge were called business cycle theory" [10].

Risk assessment play a special role for business cycle studies [21-24]. Risk measurements, concepts, techniques and tools [25] and relations to macro modeling [26] present only top slice of risk management studies. Global risk defines ground for financial policy and market risk management [26-29]. "When we speak of systemic risk, we mean the risk of a sudden, usually unexpected, disruption of information flows in financial markets that prevents them from channeling funds to those who have the most productive profit opportunities" [30]. Risk affect macroeconomic and finance development and stability [30-33] and pricing models [34].

Actually current business cycle models and description of impact of risk properties on business cycle fluctuations follow general economic equilibrium framework [7,13,23,35-38]. Endogenous business cycle models within general equilibrium framework [39-43] and relations between business cycles and risk counts hundreds of publications [21; 22; 33; 44]. "Real business cycle models view aggregate economic variables as the outcomes of the decisions made by many individual agents acting to maximize their utility subject to production possibilities and resource constraints. More explicitly, real business cycle models ask the question: How do rational maximizing individuals respond over time to changes in the economic environment and what implications do those responses have for the equilibrium outcomes of aggregate variables?" [8]. Due to [13]: "According to the standard real business cycle (RBC) approach, the competitive equilibrium of the market economy achieves resource allocation that maximizes the representative household's expected utility given the constraints on resources.

Although the RBC approach has often been criticized for its abstraction from firm and household heterogeneity, these charges are incorrect. Instead, it would be more accurate to view the RBC framework as one with heterogeneous firms and households all playing a part in the social division of labor under an ideal market mechanism. The real business cycle theory is a business cycle application of the Arrow-Debreu model, which is the standard general equilibrium theory of market economies." However, complexity and variability of business cycle properties along with development of economic processes requires different approaches and approximations. It seems unbelievable that such complex phenomena can be described by general equilibrium theory only [7,10, 45-52].

In this paper we present the business cycle model without using general equilibrium framework and assumptions as well as expectations and decisions models [53-55]. General Occam's razor principle [56] states: "Entities are not to be multiplied beyond necessity". In other words: the less initial assumptions – the better.

We don't state how real economy and markets should function – according to general equilibrium or not. Instead we suggest Econophysics approach based on assumption that it is possible to measure sufficient econometric data to assess agents risk and these assessments are adequate to real economic processes. We develop the business cycle model on assumptions that econometrics can provide sufficient data to assess risk for almost all agents of entire economics and estimate values of economic and financial transactions between agents. These "simple" assumptions hide a lot of difficulties and problems. Up now there are no econometric data sufficient for economic modelling according to our approach and our model remains as pure theory. Nevertheless we think interesting to describe macroeconomics and the business cycle within assumption that econometrics can provide required economic measurements.

Let's assume that it is possible to assess risk ratings for all agents and let's regard agents risk $x$ ratings as their coordinates $x$. Let's distribute agents by their risk ratings as coordinates on economic space [57-60]. We develop agent-based economic model but in a manner completely different from [61,62]. We do not specify particular risk under consideration and regard risk as any common economic or financial risk that can influence economic processes. All extensive (additive) macro variables are composed by aggregation of corresponding extensive variables of agents. For example, macro Investment $I(t)$ equals sum of Investment (without doubling) of all agents. Credits $C(t)$ of entire economics equal sum of Credits provided by all agents. Dynamics of extensive macro variables is determined by evolution of agents extensive economic and financial variables. Actually, transactions between agents change their variables. For example transactions that describe Credits from agent *A* to agent *B* change total Credits provided by agent

*A* and total Loans received by agent *B*. Description of transactions between agents allows model evolution of macro variables and, as we show below, can model business cycle fluctuations of macro variables.

Description of transactions between all separate agents of the economy is a very complex problem. To simplify it let's replace description of transactions between agents at points *x* and *y* by description of transactions between points *x* and *y* on economic space. To do that let's aggregate extensive variables of agents near risk point *x* in a unit volume with scales large to compare with risk scales of separate agents but small to compare with risk scales of entire economy. Scales of economy are defined by minimum or most secure risk grades and maximum or most risky grades for each particular risk. Such *roughening* of risk scales allows neglect granularity of separate agents and describe transactions between agents at points *x* and *y* as certain economic "*transaction fluids*". Such simplification is *alike to* transition from description of multi particle system in physics from kinetic to hydrodynamic approximation. This paper describes business cycles of macro variables governed by macro transactions between points *x* and *y* on economic space.

Let's explain economic ground of our approach to description of business cycle. At first let's remind that we don't use general equilibrium framework and assumptions, behavioral motivations, expectations, rational choice and decisions to describe business cycle. We don't study WHY business cycle happens. We only describe WHAT happens. We assert that appropriate **econometric observations** of agents variables, transactions between agents, agents risk assessment are sufficient to describe state and evolution of business cycle. We describe business cycle fluctuations as consequences of interactions between different macro transactions that we model by economic equations (see below (4.1-4.2) and (5.1.1-5.3)). To describe WHAT happens we propose that econometrics deliver sufficient data for risk assessments of all economic agents. We use agents risk ratings as coordinates on economic space. That distributes agents by their risk coordinates and allows aggregate variables of agents with same risk coordinates. Aggregates of extensive variables of agents with risk coordinate *x* define economic variables as function of time and coordinate *x*. For example, sum of Credits $C_j(t,x)$ of all agents *j=1,...* with risk coordinate *x* defines Credits $C(t,x)$ as function of *t* and *x*. Credits $C_j(t,x)$ of agent *j* are determined by Credits transactions $c_{jk}(t,x,y)$ between Creditor *j* at point *x* and Borrower *k* at point *y*. Aggregates off all Credit transactions from Creditors at *x* to Borrowers at *y* define macro Credit transactions $CL(t,x,y)$ as function of time and coordinates *x* and *y* on economic space. As we show below, evolution of Credits transactions $CL(t,x,y)$ define evolution of Credits $C(t,x)$ as function of *t* and *x* and Loans $L(t,y)$ as function of *t* and *y*. Total Credits $C(t)$ in economy equal sum of Credits of all agents in economy and that equals integral of Credits $C(t,x)$ by $dx$ on economic space. Distribution of Credits $C(t,x)$ as function of $x$ allows define mean Credits risk $X_C(t)$ (3.7.3; 3.7.5) as mean risk coordinates $x$ weighted by Credits $C(t,x)$ on economic space. Mean Credits risk $X_C(t)$ can be treated *alike to* center $X_C(t)$ of mass of a body with total Credit mass $C(t)$. Mean Credits risk $X_C(t)$ is not a constant. $X_C(t)$ changes due to variation of Creditors risks and changes of Credits provided by Creditors that are caused by endogenous economic and financial processes. Borders of economic domain (1) on economic space reduce motion of mean Credits risk $X_C(t)$. Thus mean Credits risk $X_C(t)$ should follow complex fluctuations on bounded economic domain (1) on economic space.

We state that mean Credits risk $X_C(t)$ fluctuates along risk axes of economic space. Fluctuations of mean Credits risk $X_C(t)$ reflect business cycle processes and are accompanied by fluctuations of total Credits $C(t)$. As we show below, motion of mean Credits risk $X_C(t)$ is governed by (see below equations (5.1.1-5.1.3; 5.2; 5.3)) complex evolution of Credits transactions $CL(t,x,y)$. Mean risk coordinates are different for different economic and financial variables and their mutual motions and interactions are very complex. Fluctuations of mean risk coordinates of different economic and financial variables reflect complex business cycle processes and accompanied by fluctuations of macro variables like Credits $C(t)$, Loans $L(t)$, Investment $I(t)$ and etc.

In [63] we derived business cycle equations under the assumption that economic and financial transactions on economic space occur between agents with same risk coordinates only. Such assumption describes *local* approximation for transactions between agents on economic space. *Local* approximation allows simplify the problem and develop economic model with *local* interactions between macro variables.

In reality economics agents with risk rating $x$ can conduct transactions – Credits, Investments and etc., to agents with any risk ratings $y$. Transactions between agents with coordinates $x$ and $y$ demonstrate economic and financial "*action-at-a-distance*" between points $x$ and $y$ on economic space. That significantly complicates macroeconomic and business cycle modeling. This paper presents a model that describe "*action-at-a-distance*" transactions between agents with any risk coordinates $x$ and $y$. We describe transactions by hydrodynamic-like economic equations on economic space. Starting with these equations we derive a system of ordinary differential equations (ODE) that describe business cycle time fluctuations of macro variables.

The rest of the paper is organized as follows. In Section 2 we present model setup and give definitions of macro transactions [60, 64]. In Section 3 we introduce a system of hydrodynamic-like economic equations on macro-transactions and discuss their economic meaning [60,64]. In Section 4 we argue economic assumptions that allow describe business cycles aggregate fluctuations. As example we study a model interactions between macro Credit transactions *CL(t,x,y)* of Loans from Creditors at point *x* to Borrowers at point *y* and macro transactions *LR(t,x,y)* of Loans Repayments from Borrowers at point *y* to Creditors at point *x*. We model these transactions by a system of economic equations and describe their evolution in a self-consistent manner. Starting with these equations we derive the system of ODE and derive simple solutions that describe business cycle fluctuations around growth trend of Credits *C(t)*. Conclusions are in Section 5.

## 2. Model Setup

In this Section we present brief definitions of economic space, explain meaning of macro variables as functions of coordinates *x* and introduce transactions between agents as functions of points *x* and *y* on economic space [57-60; 64].

Let's call the space that map agents by their risk ratings *x* as economic space. Risk ratings take values of risk grades and up now are defined by rating companies as Moody's, Fitch, S&P [65-67] as *AAA, A, BB, C* and etc. Let's regard risk grades as points $x_1, ...x_m$ of discrete space. Use of risks ratings allow distribute agents over points $x_1, ...x_m$ on discrete space. Macroeconomics and finance are under action of numerous risks. Ratings of single risk distribute agents over points of one-dimensional discrete space. Assessments of two or three risks allow distribute agents on economic space with dimension two or three. It is obvious that number of risk grades, number of points *AAA, A, BB, C...* is determined by methodology of risk assessment. Let's assume that risk assessment methodology can be generalized to make risk grades continuous so, they fill certain interval *(0,X)* on space *R*. Let's take that point *0* indicates most secure agents and point *X* denote most risky agents. Value of most risky grade *X* always can be set as *X=1* but we use *X* notation for convenience. Let's assume that risk assessments of *n* risks define agent coordinates on space $R^n$ and risk grades of *n* risks fill certain rectangle on space $R^n$.

Up now rating agencies provide risk assessments for global banks and international corporations. Let's propose now that it is possible assess risk ratings for all agents of entire economics - for global banks and corporations and for small companies and even households. It is obvious that such assumptions require a lot of additional econometric and statistical data that are absent now. Nevertheless let's propose that our assumptions are fulfilled and rating agencies are able evaluate risk assessments for all agents of entire economics.

As risk grades are continuous hence agents ratings of *n* risks fill economic domain

$$0 \leq x_i \leq X_i \; ; i = 1, \ldots n \tag{1}$$

on space $R^n$. As we mentioned above, risk grades $X_i$ always can be set as $X_i=1$. Below we study economic and financial transactions and develop business cycle model for economics that is under the action of *n* risks on economic space $R^n$. Let's assume that statistics and econometrics can provide sufficient data required for risk assessment and sufficient data to define economic and financial variables of each agent. These assumptions require significant development of current econometrics and statistics. Up now there are no econometric data required for assessment risk ratings for all economic agents but we hope that quality, accuracy and granularity of current U.S. National Income and Product Accounts system [68] gives us confidence that all econometric problems can be solved.

Agents change their extensive economic and financial variables engaging transactions with other agents. For example agent *A* can provide Credits to agent *B*. This transaction between agents will change Credits provided by agent A and Loans received by agent B. Each transaction takes certain time *dt* and we consider any transactions as rate or speed of change of corresponding variables. For example Credit transactions from agent *A* at moment *t* during time term *dt* define rate of change of total Credits provided by agent *A* till moment *t*. Let's call extensive economic or financial variables of two agents as *mutual* if output of one becomes an input of the other. For example, Credits as output of Creditors are *mutual* to Loans as input of Borrowers. Any exchange between agents by *mutual* variables is carried out by corresponding transaction. Any agent at point *x* may carry out transactions with agent at any point *y* on economic space. Different transactions define evolution of different couples of m*utual* variables. Macroeconomics as multi-agent system can be described alike to some "*economic gas*" and transactions between agents describe interactions between agent that are alike to "economic particles". For brevity and convenience let's further call economic agents as "economic particles" or e-particles and economic space as e-space. Now let's present above considerations in a more formal manner.

*2.1. Transactions between e-particles*

Let's treat Credit transactions *CL* that provide Loans from Creditors to Borrowers as example and let's follow [60; 64]. Let's denote Credit transactions $cl_{1,2}(t,\boldsymbol{x},\boldsymbol{y})$ as output of Credits $C_{out}(1,\boldsymbol{x})$ from e-particle *1* at point *x* to e-particle *2* at point *y* and equal input of Loans $L_{in}(2,\boldsymbol{y})$ of e-particle *2* from e-particle *1* at moment *t* during time term *dt*. Transaction $cl_{1,2}(t,\boldsymbol{x},\boldsymbol{y})$ describes speed of change of Credits $C_{out}(1,\boldsymbol{x})$ of e-particle *1* at point *x* due to exchange with e-particle *2* at point *y* and speed of change of Loans $L_{in}(2,\boldsymbol{y})$ of e-particle *2* at point *y* due to exchange with e-particle *1*. Credits $C_{out}(1,\boldsymbol{x})$ of e-particle *1* at point $\boldsymbol{x}$ change due to transactions $cl_{1,i}(\boldsymbol{x},\boldsymbol{y})$ with all e-particles *i* at point $\boldsymbol{y}$ during time term *dt*:

$$dC_{out}(1,\boldsymbol{x}) = \sum_i cl_{1,i}(t,\boldsymbol{x},\boldsymbol{y})\,dt$$

and vice versa

$$dL_{in}(2,\boldsymbol{y}) = \sum_i cl_{i,2}(t,\boldsymbol{x},\boldsymbol{y})\,dt$$

Thus Credit transactions $cl_{1,2}(t,\boldsymbol{x},\boldsymbol{y})$ describe Credits from e-particle *1* at point $\boldsymbol{x}$ to e-particle *2* at point $\boldsymbol{y}$. $L_{in}(2)$ equals Loans received by e-particle *2* and $C_{out}(1)$ equals Credits issued by e-particle *1* during *dt*. Sum of transactions over all input e-particles equals speed of change of output Credits $C_{out}(1)$ of e-particle *1*. Let's state that all extensive economic or financial variables can be allocated as pairs of *mutual* variables or can be describes by *mutual* variables. Otherwise there should be extensive macro variables that don't depend on any economic or financial transactions, don't depend on Markets, Investment, Credit, Buy-Sell transactions at all. It seems impossible and thus we take that transactions describe dynamics of all extensive economic and financial variables of e-particles and hence determine evolution of all extensive macroeconomic and financial variables.

*2.2 Macro transactions between points on economic space*

Let's assume that transactions between e-particles at point $\boldsymbol{x}$ and e-particles at point $\boldsymbol{y}$ are determined by exchange of *mutual* variables Credits and Loans, Buy and Sell, and etc. Different transactions describe exchange by different *mutual* variables. For example *Buy-Sell (bs)* transactions with particular Commodities, Assets, Securities and etc. at time *t* describe a case when e-particle *1* at point $\boldsymbol{x}$ during time *dt* Buy (input) amount *bs* from e-particle *2* at point $\boldsymbol{y}$ and e-particle *2* at point $\boldsymbol{y}$ at time *t* during time *dt* Sell (output) of amount *bs* to e-particle *1* at point $\boldsymbol{x}$. Further let's use Credits transaction as example to derive economic equations that describe dynamics of transactions.

Definition of macro transactions allows neglect granularity of e-particles. Main idea: let's rougher description of transactions between separate e-particles by description of transactions associated with points of e-space. In other words, let's increase e-space scales so that macro scales don't distinguish separate e-particles and transactions between them but describe aggregates of transactions between all e-particles in each large risk scale. Such a roughening is already used in economics. For example aggregation (without doubling) of all Credit transactions between agents of entire economics define macro Credit *C(t)* (see 3.6.2) provided in macroeconomics at moment *t* and equal macro Loans *L(t)* received in macroeconomics at moment *t*. Thus we have model of transactions between separate agents at points *x* and *y* on e-space that gives too detailed picture. As well current description of macroeconomic variables like macro Credits *C(t)* as functions of time only is too general as it aggregates all transactions between all agents of entire economics. We propose intermediate description of economy that aggregate transactions between agents that belong to each domain around points *x* and *y*. Such approximation neglect granularity of separate e-particles but allows take into account distribution of transactions on e-space. Such approach is similar to transition from description of kinetic multi-particle system to description to hydrodynamic approximation in physics [69-71]. For example, let's define Credit transaction *CL(t,z=(x,y))* at point *z=(x,y)* as aggregate of all Credits from e-particles at point *x* to e-particles at point *y*. As points *x* and *y* belong to *n*-dimensional e-space $R^n$ then point *z=(x,y)* can be treated as point of *2n*-dimensional e-space $R^{2n}$. Such roughening of transactions between e-particles permit describe them as "*transaction fluids*" on e-space. For example Credit transactions between e-particles defines Credit "*transaction fluids*" *CL(t,z)*, Investment transactions define Investment "*fluid*" *I(t,z)*, Buy-Sell transactions with particular commodity, define Buy-Sell "*fluid*" *BS(t,z)* for this particular commodity. Value of Credit *CL(t,z),* Investment *I(t,z),* Buy-Sell *BS(t,z)* transactions play role of corresponding densities of "*transaction fluids*" alike to mass density of physical fluid (see 3.1; 3.4; 3.5). Velocity of "*transaction fluid*" is determined as aggregates of velocities of agents that involved in transactions from point *x* to point *y* *(3.2-3.5.1)*. For example, velocities of Credit transactions fluid *CL(t,z=(x,y))* along axes $x=(x_1,...x_n)$ are determined by velocities of Creditors and along axes $y=(y_1,...y_n)$ are determined by velocities of Borrowers. Evolution of such Credit "*transaction fluids*" can be described by economic equations (4.1-4.2) [57-60; 63; 64;]. Meaning of these equations is simple: economic equations (4.1) describe balance between left and right sides. Left side of equations (4.1) describes change of Credit density *CL(t,z)* in a unit volume on *2n*-dimensional e-space. Credit density *CL(t,z)* in a unit volume can change due to change in time as $\partial CL(t,z)/\partial t$ and due to flux *CL(t,z)v(t,z)* of Credit density through surface of a unit volume. According to Divergence Theorem [72] surface integral of flux through surface of a unit volume equals volume integral of divergence and hence we obtain left side of equations (4.1). Here *v(t,z)* – velocity of Credit "*transaction fluids*" by defined (3.1 - 3.5.1). Right side describes action of other transactions on evolution of Credit "*transaction fluids*" *CL(t,z)*. These equations reflect economic properties and relations between different transactions and thus have economic nature. Meanwhile form of these equations is *alike to* hydrodynamic equations in physics of fluids and thus we may call them economic *hydrodynamic-like* equations. We underline that equations (4.1-4.2) have only formal resemblance with hydrodynamic equations as properties of *"transaction fluids"* have nothing common with physical fluids. Below we present above considerations in a more formal way.

Let's assume that e-particles on e-space $R^n$ at moment $t$ are described by coordinates $\boldsymbol{x}=(x_1,...x_n)$ and velocities $\boldsymbol{v}=(v_1,...v_n)$. Velocities $\boldsymbol{v}=(v_1,...v_n)$ describe change of e-particles risk ratings during time $dt$. Let's assume that at moment $t$ there are $N(\boldsymbol{x})$ e-particles at point $\boldsymbol{x}$ and $N(\boldsymbol{y})$ e-particles at point $\boldsymbol{y}$. Let's state that at moment $t$ e-particle $i$ at point $\boldsymbol{x}$ provide Credit $cl_{i,j}(\boldsymbol{x},\boldsymbol{y})$ to e-particle $j$ at point $\boldsymbol{y}$. Let's take Credit transactions $cl(\boldsymbol{x},\boldsymbol{y})$ between $\boldsymbol{x}$ and $\boldsymbol{y}$ as:

$$cl(t,\boldsymbol{x},\boldsymbol{y}) = \sum_{ij} cl_{ij}(t,\boldsymbol{x},\boldsymbol{y}); \quad i=1,...N(\boldsymbol{x}); i=1,...N(\boldsymbol{y}) \qquad (2.1)$$

$cl(t,\boldsymbol{x},\boldsymbol{y})$ equals growth of Credits provided by all e-particles at point $\boldsymbol{x}$ to all e-particles at point $\boldsymbol{y}$ at moment $t$ and equals rise of Loans received by all e-particles at point $\boldsymbol{y}$ from all e-particles at point $\boldsymbol{x}$ at moment $t$ during time $dt$. Transactions (2.1) between two points on e-space are random due to random number of e-particles at points $\boldsymbol{x}$ and $\boldsymbol{y}$ and random value of transactions between them. Evolution of Credit transaction $cl(t,\boldsymbol{x},\boldsymbol{y})$ depends on velocities $\boldsymbol{v}=(\boldsymbol{v}_x, \boldsymbol{v}_y)$ that describe change of risk ratings of e-particles involved in transactions at points $\boldsymbol{x}$ and $\boldsymbol{y}$. Such a treatment has *parallels* to definition of fluid velocity in hydrodynamics: motion of physical particles defines velocity of fluid [69;71]. Averaging procedure can be applied to additive variables only. Velocities of e-particles are not additive variables. To use averaging procedure let's introduce *additive* variables - transaction "impulses" $\boldsymbol{p} =(\boldsymbol{p}_x, \boldsymbol{p}_y)$ *alike to* impulses in physics [60;64]:

$$\boldsymbol{p}_x = \sum_{i,j} cl_{ij} \cdot \boldsymbol{v}_{xi}; \quad i=1,...N(\boldsymbol{x}); j=1,...N(\boldsymbol{y}) \qquad (2.2)$$

$$\boldsymbol{p}_y = \sum_{i,j} cl_{ij} \cdot \boldsymbol{v}_{yj}; \quad i=1,...N(\boldsymbol{x}); j=1,...N(\boldsymbol{y}) \qquad (2.3)$$

Here $\boldsymbol{v}_{xi}=(v_{1i},...v_{ni})$ – velocities of e-particles at point $\boldsymbol{x}$ and $\boldsymbol{v}_{yj}=(v_{1j},...v_{nj})$ – velocities of e-particles at point $\boldsymbol{y}$. Transactions *impulses* $\boldsymbol{p}_x$ and $\boldsymbol{p}_y$ are additive and admit averaging procedure by probability distribution. Transactions *impulses* $p_{Xi}$ and $p_{Yi}$, $i=1,..n$ describe flow of *"transaction fluid"* $cl(t,z=(\boldsymbol{x},\boldsymbol{y}))$ through unit surface in the direction of risks $x_i$ for Creditors and in the direction of $y_i$ for Borrowers. Credit transactions $cl(t,\boldsymbol{x},\boldsymbol{y})$ (2.1) and transactions "impulses" $\boldsymbol{p}_x$ and $\boldsymbol{p}_y$ (2.2, 2.3) take random values due to random value of transactions and motion of e-particles. To obtain regular mean impulses [60;64] let's average (2.1-2.3) by probability distribution function $f=f(t,z=(\boldsymbol{x},\boldsymbol{y}); cl, \boldsymbol{p}=(p_X,p_Y);N(\boldsymbol{x}),N(\boldsymbol{y}))$ on $2n$-dimensional e-space $R^{2n}$ that determine probability to observe Credit transactions with value $cl$ at point $z=(\boldsymbol{x}, \boldsymbol{y})$ between $N(\boldsymbol{x})$ e-particles at point $\boldsymbol{x}$ and $N(\boldsymbol{y})$ e-particles at point $\boldsymbol{y}$ with economic impulses $\boldsymbol{p} =(\boldsymbol{p}_x, \boldsymbol{p}_y)$ at time $t$. Averaging of Credit transactions and their transaction "impulses" by distribution function $f$ determine *"transaction fluid"* $CL(t,z)$ as functions of $z=(\boldsymbol{x},\boldsymbol{y})$. We do not argue here any properties of such a distribution function *f* but propose that it can be determined. Mean Credit macro transactions *CL(z=(x,y))* and "impulses" *P=(P_x,P_y)* take form:

$$CL(t, z = (x, y)) = \sum_{N(x);N(y)} \int cl \; f(t, x, y; cl, p_x, p_y; N(x), N(y)) dcl \, dp_x \, dp_y \quad (3.1)$$

$$P_x(t, z = (x, y)) = \sum_{N(x);N(y)} \int p_x \; f(t, x, y; cl, p_x, p_y; N(x), N(y)) dcl \, dp_x \, dp_y \quad (3.2)$$

$$P_y(t, z = (x, y)) = \sum_{N(x);N(y)} \int p_y \; f(t, x, y; cl, p_x, p_y; N(x), N(y)) dcl \, dp_x \, dp_y \quad (3.3)$$

That defines e-space velocity *v(t,z=(x,y))=(v_x(t,z),v_y(t,z))* of macro transaction *CL(t, z)* as:

$$P_x(t, z) = CL(t, z) v_x(t, z) \quad (3.4)$$

$$P_y(t, z) = CL(t, z) v_y(t, z) \quad (3.5)$$

$$P(t, z) = \big(P_x(t, z); P_y(t, z)\big); \; v(t, z) = \big(v_x(t, z); v_y(t, z)\big) \quad (3.5.1)$$

Let's repeat that macro Credit transactions *CL(z=(x,y))* describe mean value of Credits from all agents at point *x* to all agents at point *y* and have meaning of density of "*transaction fluids*". Impulses *P=(P_x,P_y)* describe flows of "*transaction fluids*" density *CL(t,z=(x,y))* alike to flows of physical fluids with velocities *v(t,z=(x,y))=(v_x(t,z),v_y(t,z))* on *2n*-dimensional e-space. Integral of Credit transactions *CL(t,x,y)* by variable *y* over e-space $R^n$ defines rate of change all of Credits *C(t,x)* from point *x* at moment *t*.

$$C(t, x) = \int dy \; CL(t, x, y) \; ; \; L(t, y) = \int dx \; CL(t, x, y) \quad (3.6.1)$$

Integral (3.6.1) also defines rate of change of all Loans *L(t,y)* received at point *y*. Integral of *CL(t,x,y)* by variables *x* and *y* on e-space describes rate of change of total Credits *C(t)* provided or total Loans *L(t)* received in macroeconomics at moment *t* during time term *dt*:

$$C(t) = \int dx \; C(t, x) = \int dx dy \; CL(t, x, y) = \int dy \; L(t, y) = L(t) \quad (3.6.2)$$

Relations (3.6.1; 3.6.2) show that macro transactions like Credit transactions *CL(t,x,y)* define evolution of Credits *C(t,x)* provided from point *x* and total Credits *C(t)* provided in economy at moment *t* and their *mutual* variables - Loans *L(t,y)* received at point *y* and total Loans *L(t)* received in macroeconomics at moment *t*.

Now let's introduce simple but important notion. Currently risk ratings are related with economic agents or particular Securities. Above we propose that it is possible estimate risk ratings of all agents of entire economics. If so for each particular macro variable let's define mean risks as follows. As example let use macro Credits and Loans variables. Let's assume that e-particle 1 (Bank 1) with risk coordinate *x* at moment *t* has issued Credits $C_1(t,x)$ and e-particle 2 (Bank 2) with risk coordinate *y* at moment *t* has issued Credits $C_2(t,y)$. Coordinates *x* and *y* define risk ratings of Bank1 (e-particle1) and Bank 2 (e-particle 2). What is risk rating – risk coordinate of group of both Banks? It is obvious that Credits of two Banks equal $C_1(t,x)$+ $C_2(t,y)$. Let's define Credits mean risk coordinates $X_{C1,2}(t)$ of such a group as:

$$\boldsymbol{X_{C1,2}}(t) = \frac{xC_1(t,\boldsymbol{x}) + yC_2(t,\boldsymbol{y})}{C_1(t,\boldsymbol{x}) + C_2(t,\boldsymbol{y})} \quad or \quad \boldsymbol{X_{C1,2}}(t)\big(C_1(t,\boldsymbol{x}) + C_2(t,\boldsymbol{y})\big) = xC_1(t,\boldsymbol{x}) + yC_2(t,\boldsymbol{y}) \quad (3.7.1)$$

Above relations (3.7.1) define Credits mean risk coordinates as average of risk coordinates of agents weighted by value of Credits they issue at time *t*. Similar relations for Loans $L_1(t,\boldsymbol{x})$ and $L_2(t,\boldsymbol{y})$ received by e-particles 1 and 2 at points $\boldsymbol{x}$ and $\boldsymbol{y}$ define Loans mean risk $X_{L1,2}(t)$ as:

$$\boldsymbol{X_{L1,2}}(t)\big(L_1(t,\boldsymbol{x}) + L_2(t,\boldsymbol{y})\big) = xL_1(t,\boldsymbol{x}) + yL_2(t,\boldsymbol{y}) \quad (3.7.2)$$

Thus different variables Credits *C(t,x)* and Loans *L(t,x)* determine different values of mean risk coordinates $X_{C1,2}(t)$ and $X_{L1,2}(t)$ respectively. Relations (3.7.1) are *alike to* center of Credits mass $X_{C1,2}(t)$ of two physical particles with mass $C_1(t,\boldsymbol{x})$ at point $\boldsymbol{x}$ and mass $C_2(t,\boldsymbol{y})$ at point $\boldsymbol{y}$. For Credits *C(t,x)* on e-space let's define Credits mean risk $X_C(t)$ similar to relations (3.7.1) as integral over economic domain (1) taking into account total Credits *C(t)* (3.6.2):

$$C(t)\boldsymbol{X_C}(t) = \int d\boldsymbol{x} \ \boldsymbol{x} \ C(t,\boldsymbol{x}) = \int d\boldsymbol{x}d\boldsymbol{y} \ \boldsymbol{x} \ CL(t,\boldsymbol{x},\boldsymbol{y}) \quad (3.7.3)$$

and mean Loan risk $X_L(t)$ as

$$L(t)\boldsymbol{X_L}(t) = \int d\boldsymbol{y} \ \boldsymbol{y} \ L(t,\boldsymbol{y}) = \int d\boldsymbol{x}d\boldsymbol{y} \ \boldsymbol{y} \ CL(t,\boldsymbol{x},\boldsymbol{y}) \quad (3.7.4)$$

Mean Credits risk $X_C(t)$ equals mean risk coordinates of total Credits *C(t)* in economy. It is *alike to* center of mass $X_C(t)$ of a body with total mass *C(t)* and mass density *C(t,x)*. Mean Risk $X_L(t)$ defines Loans mean risk coordinates of total Loans *L(t)* in economy. We introduced notions of mean risks in [63] as mean risk or mean coordinates of e-particles averaged by particular distribution. Let's repeat - mean Credit risk $X_C(t)$ equals mean risk coordinates of e-particles averaged by Credits distribution *C(t,x)* and mean Loan risk $X_L(t)$ equals mean risk coordinates of e-particles averaged by Loans distribution *L(t,x)*. Different economic variables as Investment *I(t,x)*, Assets *A(t,x)* and etc. define different values of their mean risks. Let's remind that all variables are determined by corresponding economic transactions due to relations (3.6.1). Credit transactions mean risk of *CL(t,z=(x,y))* define mean risk of mutual variables for *z=(x,y)* as:

$$\{C(t)\boldsymbol{X_C}(t) \ ; \ L(t)\boldsymbol{X_L}(t)\} = \int d\boldsymbol{z} \ CL(t,\boldsymbol{z} = (\boldsymbol{x},\boldsymbol{y})) =$$

$$\{\int d\boldsymbol{x}d\boldsymbol{y} \ \boldsymbol{x} \ CL(t,\boldsymbol{x},\boldsymbol{y}) \ ; \ \int d\boldsymbol{x}d\boldsymbol{y} \ \boldsymbol{y} \ CL(t,\boldsymbol{x},\boldsymbol{y})\} \quad (3.7.5)$$

Relations (3.7.5) show that macro transactions like Credits transactions *CL(t,x,y)* determine evolution of Credits mean risks $X_C(t)$ and Loans mean risks $X_L(t)$. The same correct for mean risks determined by other macro transactions.

Why we attract attention to definition of mean risks? We propose that evolutions of mean risks for different macro variables describe business cycle fluctuations of these variables. Let's take Credits *C(t,x)* as example. Credits mean risk $X_C(t)$ is not a constant. It changes due to change of coordinates $\boldsymbol{x}$ and amount of Credits provided by e-particles (agents). Growth of risks of e-particles can increase and decline of Credits risk can decrees Credits mean risk $X_C(t)$.

Economy is defined on economic domain (1) on e-space. Risk ratings of e-particles (economic agents) are bounded (1) by minimum or most safe grades and maximum or most risky grades. Thus Credits mean risk $X_C(t)$ as well as mean risks of any macro variable can't grow up or diminish steadily along each risk axes as their values are bounded on economic domain (1). Values of mean risks and value of Credits mean risk $X_C(t)$ in particular along each risk axes should oscillate from certain minimum to maximum values and these fluctuations can be very complex.

We propose that business cycles correspond to fluctuations of mean risks of macro variables. Growth of Credits mean risk $X_C(t)$ can correspond with growth of total Credits $C(t)$ provided in economy and decline of Credits mean risk can correspond with total Credits contraction. Relations between Credits mean risk $X_C(t)$ and value of total Credits $C(t)$ are much more complex but we repeat main statement: business cycles can be treated as fluctuations of mean risks for different macro variables.

As we show in (3.7.5) Credits macro transaction $CL(t,x,y)$ determine Credits $X_C(t)$ and Loans $X_L(t)$ mean risks. Below in Sec. 3, Sec.4 and in Appendix we describe model dynamics of Credits transaction $CL(t,x,y)$ on e-space by economic equations (5.1.1-5.1.3; 5.2; 5.3). Starting with these equations we derive the system of ODE (A.4; A.8.4-7; A.9.6-7) that describe business cycle fluctuations of total Credits $C(t)$ provided in economy and total Loans $L(t)$ received in economy as consequences of fluctuations of Credits and Loans mean risks $X_C(t)$ and $X_L(t)$.

Due to (3.6.1) total value of macro Credits $MC(t,x)$ provided from point $x$ up to moment $t$ equal:

$$\frac{\partial}{\partial t} MC(t, x) = C(t, x) \; ; \; MC(t, x) = MC(0, x) + \int_0^t d\tau \int dy \; CL(\tau, x, y) \qquad (3.8)$$

Total value of macro Loans $ML(t,y)$ received at point $y$ up to moment $t$

$$\frac{\partial}{\partial t} ML(t, x) = L(t, x) \; ; \; ML(t, y) = ML(0, y) + \int_0^t d\tau \int dx \; CL(\tau, x, y) \qquad (3.9)$$

Here $MC(0,x)$ define initial values of total macro Credits issued from point $x$ on e-space. Relations similar to (3.6.1 - 3.9) define evolutions and fluctuations of all extensive economic and financial variables determined by macro transactions. Aggregate macro Credits $MC(t)$ issued in entire economics equal (see 3.6.2; 3.8):

$$MC(t) = MC(0) + \int_0^t d\tau \int dxdy \; CL(\tau, x, y) = MC(0) + \int_0^t d\tau \; C(\tau) \qquad (3.10)$$

Thus to describe Business or Credit cycle fluctuations of $MC(t)$ one should describe rate of change of total Credits $C(t)$ and Credit transactions $CL(t,x,y)$ (3.11):

$$\frac{d}{dt} MC(t) = C(t) = \int dxdy \; CL(t, x, y) \qquad (3.11)$$

Oscillations of rate of change of Credits *C(t)* define business cycle fluctuations of aggregate macro Credits *MC(t)*. Relations (3.1-3.11) establish basis for modeling cycle fluctuations of economic and financial variables via description of macro transaction dynamics. Below we derive economic equations to describe evolution of Credit *"transaction fluid" CL(t,x,y)*.

## 3. Equations on macro transactions

Macro transactions between points *x* and *y* on e-space determine evolution of macro variables (3.6.1 – 3.11). Let's explain derivation of economic equations on macro transactions according to [60;64]. Let's explain factors that cause change of macro "transaction fluids". Value of Credit transactions *CL(t, z=(x,y))* (3.1) play role of transaction fluid density. Credit fluid density *CL(t, z)* in a unit volume *dV* at point *z=(x,y)* can change due to two factors. First factor describes change of *CL(t,z)* in time as $\partial CL/\partial t$. Second factor describes change of *CL(t,z)* in a unit volume *dV* due to flux of transactions flow *CLv* through surface of a unit volume. Divergence theorem [72] state that flux through surface of a unit volume equals volume integral of divergence. Thus total change of transaction *CL(t,z)* in a unit volume *dV* equals

$$\frac{\partial CL}{\partial t} + \nabla \cdot (vCL)$$

Here *v=($v_X$,$v_Y$)* – velocity of transaction *CL(t,z=(x,y))* on *2n*-dimension e-space $R^{2n}$ determined by (3.4-3.5), bold letters *x, y, z, P, $Q_2$* mean vectors, roman *t, CL* mean scalars and divergence equals:

$$\nabla \cdot (vCL) = \sum_{i=1,\ldots n} \frac{\partial}{\partial x_i}\left(v_{xi}(t,x,y)CL(t,x,y)\right) + \sum_{i=1,\ldots n} \frac{\partial}{\partial y_i}\left(v_{yi}(t,x,y)CL(t,x,y)\right)$$

Now let's mention that such change of transactions *CL(t,z)* can be induced by action of other transactions. Let's denote action of other transactions on *CL(t,z)* that cause change of Credit transactions *CL(t,z=(x,y))* as factor $Q_1$. Then equation takes form:

$$\frac{\partial CL}{\partial t} + \nabla \cdot (vCL) = Q_1 \qquad (4.1)$$

Thus left side (4.1) describes how *CL(t,z)* can change in a unit volume – due to change in time and due to flux through surface of a unit volume. Right side describes action of other transactions. Equation (4.1) is a simple balance of factors that can change of *CL(t,z)*. The same reasons define equation on transaction impulses *P(t,z)=($P_x$(t,z) $P_y$(t,z))* determined by (3.2-3.3) as:

$$\frac{\partial P}{\partial t} + \nabla \cdot (vP) = Q_2 \qquad (4.2)$$

Thus left side of (4.2) describes change of transaction impulses *P(t,z)=($P_x$(t,z), $P_y$(t,z))* due to change in time $\partial P/\partial t$ and due to flux *vP* through surface of unit volume that equal divergence $\nabla \cdot (\boldsymbol{v P})$ and right hand side $Q_2$ describe action of other transactions on evolution of transaction impulses. Equations (4.1; 4.2) present a balance between changes of transactions *CL(t,z)* and their impulses *P(t,z)* in the left side and action of other transactions that can induce these changes. To describe a particular economic model via equations (4.1; 4.2) let's determine direct form of right hand side $Q_1$ and $Q_2$. In hydrodynamics [70] equations alike to (4.1; 4.2) are called as Continuity Equations and Equations of Motion and for convenience let's further use same notions. However we underline that hydrodynamic equations and laws that describe physics of fluids [70] have nothing common with equations (4.1; 4.2) and economics. We only use similarities between fluid dynamics and economics to develop reasonable model of business cycles according to economic laws.

Macro transactions *CL(t,z)* and their impulses *P(t,z)* can depend on various transactions. Above we propose that all extensive macro variables should be determined by macro transactions or depend on variables that are described by macro transactions. That means that macro transactions should depend on other transactions only. Let's study simplest case and assume that factors $Q_1$ and $Q_2$ in (4.1-4.2) depend on only one transaction Loan-Repayment *LR(t,z)* that describe payout on Credits by Borrowers from point *y* to Creditors at point *x*. To simplify the problem let's assume that that factors $Q_1$ and $Q_2$ that define equations on Loan-Repayment transactions *LR(t,z)* and their impulses depend on Credit transactions *CL(t,z)* only. Our assumptions propose that Credits from point *x* to point *y* are provided at time *t* due to Loan-Repayments received at same time *t* and vice versa. Such assumptions simplify mutual dependence between Credit transactions *CL(t,z)* and Loan-Repayment *LR(t,z)* and allow develop simple description of business cycle fluctuations of macro Credits *C(t)* issued at time *t*.

**4 How macro transactions describe business cycles**

In [63] we proposed that agents perform only *local* economic or financial transactions with agents at same point *x*. Such simplification describes interactions between macro variables at point *x* by *local* operators. This paper describes a model of transactions that can occur between agents at arbitrary points *x* and *y*. Such transactions describe *non-local* economic and financial "*action-at-a-distance*" between agents at points *x* and *y* on e-space $R^n$. Below we describe a model of business cycle fluctuations determined by *non-local* Credit *CL(t,z)* and Loan-Repayment *LR(t,z)* transactions. To describe evolution of transactions *CL(t,z)* and *LR(t,z)* let's take into account their mutual interactions on e-space. Let's assume that *CL(t,z)* at point *z=(x,y)* on e-space $R^{2n}$ depend on Loan-Repayment *LR(t,z)* transactions and their impulses *L(t,z)* only and vice versa. To define factors $Q_1$ and $Q_2$ let's simplify the problem and assume that $Q_{11}$ for Continuity Equation (4.1) on macro transactions *CL(t,z)* at point *(t,z)* is proportional to scalar product of vector *z* and Loan-Repayment impulse *D(t,z)*

$$Q_{11} = a\, \mathbf{z} \cdot \mathbf{D}(t, \mathbf{z}) = a(\, \mathbf{x} \cdot \mathbf{D_x}(t, \mathbf{z}) + \mathbf{y} \cdot \mathbf{D_y}(t, \mathbf{z}))$$

Loan-Repayment impulse *D(t,z)* and velocity *u(t,z)* are determined similar to (3.1-3.5.1). Scalar product is a simple linear operator and we use it to demonstrate capabilities of our approach. Let's assume that same relations define factor $Q_{12}$ for Continuity Equation (4.1) on Loan-Repayment *LR(t,z)* macro transactions:

$$Q_{12} = b\, \mathbf{z} \cdot \mathbf{P}(t, \mathbf{z}) = b(\mathbf{x} \cdot \mathbf{P_x}(t, \mathbf{z}) + \mathbf{y} \cdot \mathbf{P_y}(t, \mathbf{z}))$$

Here *a* and *b* – const and Continuity Equations on transactions *CL(t,z)* and *LR(t,z)* take form:

$$\frac{\partial CL}{\partial t} + \nabla \cdot (\mathbf{v} CL) = Q_{11} = a\, \mathbf{z} \cdot \mathbf{D}(t, \mathbf{z}) = a\, (\mathbf{x} \cdot \mathbf{D_x}(t, \mathbf{z}) + \mathbf{y} \cdot \mathbf{D_y}(t, \mathbf{z}))\quad (5.1.1)$$

$$\frac{\partial LR}{\partial t} + \nabla \cdot (\mathbf{u} LR) = Q_{12} = b\, \mathbf{z} \cdot \mathbf{P}(t, \mathbf{z}) = b\, \left(\mathbf{x} \cdot \mathbf{P_x}(t, \mathbf{z}) + \mathbf{y} \cdot \mathbf{P_y}(t, \mathbf{z})\right)\quad (5.1.2)$$

$$\mathbf{P}(t, \mathbf{z}) = \mathbf{v}(t, \mathbf{z}) CL(t, \mathbf{z})\; ;\; \mathbf{D}(t, \mathbf{z}) = \mathbf{u}(t, \mathbf{z}) LR(t, \mathbf{z})\quad (5.1.3)$$

Economic meaning of (5.1.1-5.1.3) is as follows. *CL(t,z)* in a unit volume at point *(t,z)* grows up if $Q_{11}$ is positive. A position vector *z* has origin at secure point *0* and points to risky point *z*. Hence for *a>0* positive value of $\mathbf{z} \cdot \mathbf{D}(t, \mathbf{x})$ models Loan-Repayment flow

$$\mathbf{D}(t, \mathbf{x}) = LR(t, \mathbf{x}) \mathbf{u}(t, \mathbf{x})$$

in risky direction *z* and that can induce growth of Credits *CL(t,z)* to risky points. As well negative value of $\mathbf{z} \cdot \mathbf{D}(t, \mathbf{x})$ models Loan-Repayment flows from risky to secure domain and that can decrease Credits *CL(t,z)* as Creditors can prefer more secure Borrowers. This model simplifies Credit modeling as it neglect time gaps between providing Credits from point *x* to point *y* and Loan-Repayment received from Borrowers at point *y* to Creditors at point *x* and neglect other factors that can impact on providing Credits. To determine $Q_{21}$ factor that define Equations of Motion (4.2) on Credit impulses *P(t,z)* let's assume that $Q_{21}$ is a linear operators on Loan-Repayment impulses *L(t,z)* and in a matrix form take form:

$$\mathbf{Q}_{21} = \widehat{\Omega} \mathbf{D}(t, \mathbf{z}) = \Omega_{ij} D_j(t, \mathbf{z})$$

Let's assume that $Q_{22}$ factor that define Equations of Motion (4.2) on Loan-Repayment impulses *L(t,z)* is similar linear operator:

$$\mathbf{Q}_{22} = \widehat{\Phi} \mathbf{P}(t, \mathbf{z}) = \Phi_{ij} P_j(t, \mathbf{z})$$

and Equations of Motion for impulses *P(t,z)* and *L(t,z)* take form:

$$\frac{\partial \mathbf{P}}{\partial t} + \nabla \cdot (\mathbf{v}\, \mathbf{P}) = \mathbf{Q}_{21} = \Omega \mathbf{D}(t, \mathbf{z}) = \Omega_{ij} D_j(t, \mathbf{z}) = \Omega_{xij} D_{xj}(t, \mathbf{z}) + \Omega_{yij} D_{yj}(t, \mathbf{z})\; (5.2)$$

$$\frac{\partial \mathbf{D}}{\partial t} + \nabla \cdot (\mathbf{u}\, \mathbf{D}) = \mathbf{Q}_{22} = \Phi \mathbf{P}(t, \mathbf{z}) = \Phi_{ij} P_j(t, \mathbf{z}) = \Phi_{xij} P_{xj}(t, \mathbf{z}) + \Phi_{yij} P_{yj}(t, \mathbf{z})\; (5.3)$$

Equations (5.2-5.3) describe simple linear mutual dependence between transaction impulses *P(t,z)* and *D(t,z)*. Meaning of equations (5.2; 5.3) can be explained is as follows. Let's mention that integral of each component of impulses *P(t,z)* or its components $P_{xi}(t,z)$ and $P_{yi}(t,z)$ along axes $x_i$ or $y_i$ over *dz* define total impulses *P(t)* and its components $P_{xi}(t)$ or $P_{yi}(t)$ along risk axis $x_i$ or $y_i$ and due to (3.4; 3.5; A.6.3.1; A.6.3.2):

$$P_{xi}(t) = \int dz \, P_{xi}(t, \mathbf{z} = (\mathbf{x}, \mathbf{y})) = \int d\mathbf{x} d\mathbf{y} \, P_{xi}(t, \mathbf{x}, \mathbf{y}) = C(t) v_{xi}(t) \quad (5.3.1)$$

$$\mathbf{P}(t) = (\mathbf{P}_C(t); \mathbf{P}_B(t)); \quad \mathbf{P}_C(t) = \mathbf{P}_x(t); \quad \mathbf{P}_B(t) = \mathbf{P}_y(t) \quad (5.3.2)$$

Total impulses *P(t)* (5.3.2) have component of Creditors impulses $P_C(t) = P_x(t)$ along axes *x* and component $P_B(t) = P_y(t)$ of Borrowers impulses along axes *y*. As we show below equations (5.2; 5.3) lead to equations (A.6.6-8) that describe fluctuations of total impulses *P(t)*. Due to (A.4.2) total impulses (5.3.1) describe fluctuations of Credits mean risk $X_{Ci}(t)$ along each risk axes $x_i$. Thus equations (5.2; 5.3) present model dynamics of Credits and Loans impulses *P(t,z)* and *D(t,z)* that cause fluctuations of Credits and Loans mean risks $X_C(t)$ and $X_L(t)$. Hence equations (5.1.1-5.1.3) and (5.2; 5.3) present model of business cycle and as we show below (A.11) describe business cycle fluctuations of total Credits *C(t)* and Loans *L(t)*. We repeat definitions of total Credits *C(t)*, total Loan-Repayment *LR(t)* and their total impulses *P(t)* and *D(t)* for convenience:

$$C(t) = \int d\mathbf{x} d\mathbf{y} \, CL(t, \mathbf{x}, \mathbf{y}) \; ; \quad LR(t) = \int d\mathbf{x} d\mathbf{y} \, LR(t, \mathbf{x}, \mathbf{y}) \quad (5.4.1)$$

$$\mathbf{P}(t) = \int d\mathbf{x} d\mathbf{y} \, \mathbf{P}(t, \mathbf{x}, \mathbf{y}) = \int d\mathbf{x} d\mathbf{y} \, CL(t, \mathbf{x}, \mathbf{y}) \, \mathbf{v}(t, \mathbf{x}, \mathbf{y}) = C(t) \, \mathbf{v}(t) \quad (5.4.2)$$

$$\mathbf{D}(t) = \int d\mathbf{x} d\mathbf{y} \, \mathbf{D}(t, \mathbf{x}, \mathbf{y}) = \int d\mathbf{x} d\mathbf{y} \, LR(t, \mathbf{x}, \mathbf{y}) \, \mathbf{u}(t, \mathbf{x}, \mathbf{y}) = LR(t) \, \mathbf{u}(t) \quad (5.4.3)$$

To describe cycle fluctuations of macro variables we derive the system of ODE (Appendix: A.4; A.8.4-7; A.9.6-7) on aggregate variables *C(t), LR(t)* starting with equations (5.1.1-5.1.3) and equations (5.2; 5.3) and present elementary solutions (A.10) for business cycle under action of a single risk. The simplest case of business cycle fluctuations of total Credits *C(t)* under action of a single risk can be derived from (A.11) with *C(j)=const, j=0,1,2,3*:

$$C(t) = C(0) + a \, [C(1) \sin \omega t + C(2) \cos \nu t + C(3) \exp \gamma t] \quad (6.1)$$

Due to (3.10; 6.1) total Credits *MC(t)* provided in economy during time term [0,t] take form:

$$MC(t) = MC(0) + \left[C(0)t + a \frac{C(3)}{\gamma} \exp \gamma t\right] + a \left[\frac{C(2)}{\nu} \sin \nu t - \frac{C(1)}{\omega} \cos \omega t\right] \quad (6.2)$$

Relations (6.1; 6.2) describe business cycle fluctuations of total Credits *C(t)*. Frequencies of business cycle fluctuations are determined by oscillations of Creditors impulses $P_x(t)$ with frequencies *ω* and oscillations of Borrowers impulses $P_y(t)$ with frequencies *ν* (Appendix, A.6.6-10; A.8.4-7; A.9.6-7). Business cycle fluctuations (6.1; 6.2) happen about exponential growth trend *exp(γt)* (Appendix, A.9.5-7) and we take coefficient *γ* =max($γ_x$, $γ_y$). Thus *γ* describes maximum growth trend induced by (A.8.6-7; A.9.1-2; A.10.1-2). Factors (A.8.6) are proportional to product of total Credits *C(t)* and square of transactions velocity $v^2(t)$ and we call them Credits "energy" because they are *alike to* kinetic energy of body with mass equals *C(t)* and square of velocity $v^2(t)$. However meaning of Credits "energy" have nothing common with meaning of energy in physics as no conservation laws are valid for this variable.

Total macro Credits *MC(t)* made during time term *[0,t]* are described by (6.2). If initial value *C(0)* is not zero then macro Credits *MC(t)* has linear and exponential growth trend and oscillations with same frequencies *ω* and *v* about these trends. Solutions (6.1) for Credit transactions *C(t)* and for Loan-Repayment transactions *LR(t)* present simplest form of Credit cycle fluctuations determined by action of single risk and simple interactions between two macro transactions (Appendix). Action of several risks can make Credit and Business cycle fluctuations more complex (A.11). If one neglect growth trend then business cycle fluctuations of Credits *C(t)* under action of *n* risks can take form (A.11):

$$C(t) = C(0) + a\sum_{i=1}^{n}[\,C_{xi}(1)\sin\omega_i t + C_{xi}(2)\cos\omega_i t + C_{yi}(3)\sin v_i t + C_{yi}(4)\cos v_i t\,] \quad (6.3)$$

In (6.3) frequencies $\omega_i$ reflect oscillations of Credit impulses ***P(t)*** along axes $x_i$, and frequencies $v_i$ along axes $y_i$, *i=1,..n* on *2n* dimensional e-space *(**x,y**)* (Appendix)

## 5. Conclusions

Business cycle fluctuations are extremely complex and their behavior is under permanent evolution due to development of entire economy. It seems impossible establish single, precise, exact description of such alive phenomena and each business cycle model should be based on definite assumptions and simplifications. Occam's razor [56] principle states that the less initial assumptions are made - the better. We don't use initial assumptions of general equilibrium framework and hence our business cycle model may be treated as reasonable alternative to mainstream economics at least.

We make no general equilibrium assumptions on state and evolution of markets, prices and consumers decisions, but make an attempt to describe business cycles on base of econometric observations and risk assessments. We propose that econometrics can provide sufficient data for risk assessments of all agents of entire economics and treat agents risk ratings as their coordinates on economic space. Assessment of two-three risks defines agents risk coordinates on economic space with dimension 2 or 3. Risk coordinates distribute economic agents over points of economic space. All extensive (additive) macroeconomic and financial variables are defined as sum (without doubling) of corresponding variables of economic agents. Evolution of agents variables is conducted by economic and financial transactions between agents. Transactions between agents describe rate of change of *mutual* variables of agents.

Macroeconomic model based on description of transactions between agents takes into account granularity of agents on economic space and has some parallels to kinetic description of multi-particle systems in physics. We propose transition from description of transactions between agents at points *x* and *y* to description of transactions between points *x* and *y* on economic space. That is alike to transition from kinetic approximation that takes into account granularity of physical particles to hydrodynamic approximation that describes systems as physical fluids and neglect granularity of physical particles [69; 70]. We underline vital distinctions between economic and physical processes and remind that we use only analogies between economics and physics and don't apply physical results to economic modeling. We collect all transactions between agents at points *x* and *y* and average them by probability distribution (3.1-3.5.1). Mean values of transactions at point *z=(x,y)* has meaning alike to "*transaction fluids*" and its evolution can be describe by economic equations (4.1-4.2). These economic equations for transactions *CL(t,z)* describe balances between left side and right side factors. Left side factors describe change of Credits transaction *CL(t,z)* in a unit volume due to time derivative in time and due to flux through surface of a unit volume. Right side factors describe possible action of other transactions on *CL(t,z)*. Motion of "*transaction fluids*" is determined by average collective velocity of agents at points *x* and *y* respectively and variations of corresponding transactions between agents (3.2-3.5). Velocity of agents on economic space is defined as change of risk ratings during time term *dt*. Agents of entire economics fill economic domain (1) on economic space that is bounded by minimum (most secure) and maximum (most risky) risk grades (1; A.1). Motion of agents as well as motion of each "*transaction fluid*" causes movement of corresponding mean risk *X(t)*. For example motion of total Credits *C(t)* that is described by Credit impulse $P_x(t)$ causes motion of Credits mean risk $X_C(t)$ (A.4.2). Motion of Credits mean risk $X_C(t)$ can't go on steadily in one direction, as it will reach secure or risky boundaries of economic domain (1). Thus Credits mean risk $X_C(t)$ should fluctuate and that should accompanied by business cycle fluctuations of total Credits *C(t)*. We propose that fluctuations of Credit mean risks $X_C(t)$ reflect Credits cycle fluctuations.

To show benefits of our approach we present a simple model of interactions between Credit transactions *CL(t,z)* and Loan-Repayment transactions *LR(t,z)*. We study a model interactions between these transactions and derive system of economic equations (5.1.1-5.1.3; 5.2-5.3) in explicit and self-consistent form. Starting with these economic equations we derive the system of ODE (A.4; A.8.4-7; A.10.1-2) that describe business cycle time fluctuations of rate of change of total Credits *C(t)* and macro Credits *MC(t)*. For simplest case of business cycle fluctuations under action of single risk we derive solutions (6.1) for total Credits *C(t)*. We outline that system of ODE (A.4; A.8.4-7; A.10.1-2) contain equations for economic factors (A.8.6-8.7; A.9.1-9.2; A.10.1-10.2) that are *alike to* kinetic energy. For example factors $ECx_i(t)$ and $ECy_i(t)$ (A.8.6) are proportional to product of total Credits $C(t)$ and square of velocity $v^2_{xi}$ and $v^2_{yi}$ along risk axes $x_i$ or $y_i$ and that is *looks like* kinetic energy of body with Credits mass $C(t)$ and square of velocity $v^2$. Nevertheless these parallels have no further development it is very interesting that description of Credit cycle fluctuations requires equations (A.9.1-9.2) on factors (A.8.8-8.9) that are *alike to* Credits "*energy*".

Our approach has certain parallels to input-output analysis [73] as its macroeconomic model is based on description of macro transactions between different industries. Meanwhile, breakdown of macroeconomics by Sectors and Industries does not define any metric space. Our macroeconomic model describe macro transactions between points **x** and **y** of metric economic space. This "small" alterity permit define macro variables and macro transactions as functions of time and coordinates **x** and **y** on economic space. Such approach uncovers hidden complexity of macroeconomic and financial processes and for sure requires usage of mathematical physics methods and equations.

Comparison of our model with observed business cycles requires a lot of econometric data that could specify risk ratings of economic agents, their economic and financial variables, economic and financial transactions between agents. Absence of sufficient econometric data up now makes our business cycle model pure theoretical. Econometric assessment of our theory requires development of risk assessment methodology that allows estimate risk ratings for continuous risk grades. Usage of macroeconomic modeling on economic space requires methods that can estimate influence of particular risk on economic evolution and selection of *n* major risks that form representation of economic space. Nevertheless no principal obstacles exist that can prevent development of econometrics in a way sufficient for modeling business cycles on economic space. We propose that our theory can help financial authorities, Central Banks and business communities to forecast and manage business cycles.

## Economic Transactions and Business Cycle Equations

Let's study transactions between agents on *n*-dimensional e-space $R^n$. We use standard notations: bold letters like ***P, v, x, y, z*** define vectors and roman *C, CL, X,…* - scalars. Vector ***z*=(*x,y*)** is defined on *2n*-dimensional e-space $R^{2n}$. Scalar product:

$$\boldsymbol{z} \cdot \boldsymbol{P} = \boldsymbol{x} \cdot \boldsymbol{P_x} + \boldsymbol{y} \cdot \boldsymbol{P_y} = \sum_{i=1,..n} x_i P_{xi} + \sum_{i=1,..n} y_i P_{yi}$$

Divergence equals:

$$\nabla \cdot (\boldsymbol{v}f) = \nabla_x \cdot (\boldsymbol{v_x}f) + \nabla_y \cdot (\boldsymbol{v_y}f) = \sum_{i=1,..n} \frac{\partial}{\partial x_i}(v_{xi}f) + \sum_{i=1,..n} \frac{\partial}{\partial y_i}(v_{yi}f)$$

$$\nabla \cdot (\boldsymbol{v}\boldsymbol{P}) = \big(\nabla \cdot (\boldsymbol{v}\boldsymbol{P_x}); \nabla \cdot (\boldsymbol{v}\boldsymbol{P_y})\big) = \big(\nabla \cdot (\boldsymbol{v}P_{xj}); \nabla \cdot (\boldsymbol{v}P_{yj})\big); j = 1,..n$$

Integral notations:

$$\int d\boldsymbol{z} = \int d\boldsymbol{x}\, d\boldsymbol{y} = \int dx_1 \dots dx_n dy_1 \dots dy_n$$

To derive a system of ODE on speed of total Credit *C(t)* and Loan-repayment *LR(t)* change let's start with equations (5.1.1). Thus Credit transactions *CL(t,**z**=(**x,y**))* are determined on *2n*-dimensional e-space and economic domain (1) define *2n*-dimensional economic area ***z*=(*x,y*)**:

$$0 \le x_i \le X_i; \; 0 \le y_i \le X_i \; i = 1, \dots n \tag{A.1}$$

Let's remind that similar to (1) values of $X_i$ can be set as $X_i=1$. To derive equations on *C(t)* (5.4.1) let's take integral by *d**z**=d**x**d**y*** of equation (5.1.1):

$$\frac{d}{dt}C(t) = \frac{d}{dt}\int d\boldsymbol{z}\, CL(t,\boldsymbol{z}) = -\int d\boldsymbol{z}\, \nabla \cdot \big(\boldsymbol{v}(t,\boldsymbol{z})CL(t,\boldsymbol{z})\big) + a\int d\boldsymbol{z}\, \boldsymbol{z} \cdot \boldsymbol{D}(t,\boldsymbol{z}) \tag{A.2.1}$$

First integral in the right side (A.2.1) equals integral of divergence over *2n* dimensional e-space and due to divergence theorem [76] equals integral of flux through surface. Thus it equals zero, as no economic or financial fluxes exist far from boundaries of economic domain (A.1).

$$\int d\boldsymbol{z}\, \nabla \cdot \big(\boldsymbol{v}(t,\boldsymbol{z})CL(t,\boldsymbol{z})\big) = 0 \tag{A.2.2}$$

Let's define *Pz(t)* and *Lz(t)* as:

$$Pz(t) = \int d\boldsymbol{z}\, \boldsymbol{P}(t,\boldsymbol{z}) \cdot \boldsymbol{z} = \int d\boldsymbol{x} d\boldsymbol{y}\, \sum_{i=1}^{n} x_i P_{xi}(t,\boldsymbol{x},\boldsymbol{y}) + \int d\boldsymbol{x} d\boldsymbol{y} \sum_{i=1}^{n} y_i P_{yi}(t,\boldsymbol{x},\boldsymbol{y}) \tag{A.3.1}$$

$$Dz(t) = \int d\boldsymbol{z}\, \boldsymbol{D}(t,\boldsymbol{z}) \cdot \boldsymbol{z} = \int d\boldsymbol{x} d\boldsymbol{y}\, \sum_{i=1}^{n} x_i D_{xi}(t,\boldsymbol{x},\boldsymbol{y}) + \int d\boldsymbol{x} d\boldsymbol{y} \sum_{i=1}^{n} y_i D_{yi}(t,\boldsymbol{x},\boldsymbol{y}) \tag{A.3.2}$$

Due to (5.1.1; 5.1.2; 5.4.1; A.2.1) equations on *C(t)* and *LR(t)* take form:

$$\frac{d}{dt}C(t) = a\, Dz(t) \quad ; \quad \frac{d}{dt}LR(t) = b\, Pz(t) \tag{A.4}$$

Equation (5.1.1) permits derive equation on Credits mean risk $X_C(t)$ and Loans mean risk $X_L(t)$ (3.7.3 - 3.7.5). Let's multiply (5.1.1) by ***z*** and take integral by *d**z**=d**x**d**y***

$$\frac{d}{dt}\int d\boldsymbol{z}\, CL(t,\boldsymbol{z})\boldsymbol{z} = -\int d\boldsymbol{z}\, \boldsymbol{z}\, \nabla \cdot \big(\boldsymbol{v}(t,\boldsymbol{z})CL(t,\boldsymbol{z})\big) + a\int d\boldsymbol{z}\, \boldsymbol{z}\,(\boldsymbol{z} \cdot \boldsymbol{D}(t,\boldsymbol{z})) \tag{A.4.1}$$

We refer [71] for derivation of complete equations on mean risk. From (A.4.1) one can obtain:

$$\frac{d}{dt}C(t)X_C(t) = \boldsymbol{P}_x(t) + a\,(XDx(t) + XDy(t)) \tag{A.4.2}$$

$$\frac{d}{dt}L(t)X_L(t) = \boldsymbol{P}_y(t) + a\,(YDx(t) + YDy(t)) \tag{A.4.3}$$

$$XDx(t) = \int dxdy\,x\,(\boldsymbol{x}\cdot\boldsymbol{D}_x(t,\boldsymbol{z})) \;;\; XDy(t) = \int dxdy\,x\,(\boldsymbol{y}\cdot\boldsymbol{D}_y(t,\boldsymbol{z}))$$

$$YDx(t) = \int dxdy\,y\,(\boldsymbol{x}\cdot\boldsymbol{D}_x(t,\boldsymbol{z})) \;;\; YDy(t) = \int dxdy\,y\,(\boldsymbol{y}\cdot\boldsymbol{D}_y(t,\boldsymbol{z}))$$

Equations on factors *XDx(t), XDy(t), XDx(t), XDy(t)* can be derived similar to [71] and for brevity we omit it here. In the absence of any interaction for *a=0* equations (A.4.2; A.4.3) show that dynamics of *C(t)X_C(t)* and *L(t)X_L(t)* depends on $\boldsymbol{P}_x(t)$ and $\boldsymbol{P}_y(t)$

$$\frac{d}{dt}C(t)X_C(t) = \boldsymbol{P}_x(t) = C(t)\boldsymbol{v}_x(t)\,;\, \frac{d}{dt}L(t)X_L(t) = \boldsymbol{P}_y(t) = L(t)\boldsymbol{v}_y(t) \tag{A.4.4}$$

Thus equations (A.6.6-6.8) that describe fluctuations of impulses $\boldsymbol{P}_x(t)$ and $\boldsymbol{P}_y(t)$ cause fluctuations of *C(t)X_C(t)* and *L(t)X_L(t)*. Interactions between transactions (A.4.2; A.4.3) for *a≠0* make these fluctuations much more complex. To avoid excess complexity here we don't derive complete system of ODE on *C(t)X_C(t)* and *L(t)X_L(t)*.

To derive equations on *Pz(t)* and *Dz(t)* let's use equations on impulses *P(t), D(t)*. Let's start with (5.3; 5.4). To simplify derivation of equations let's take matrix operators in equations (5.3; 5.4) in simplest diagonal form *( i=1,..n )*:

$$\Phi_{ij} = (\Phi_{xij};\Phi_{yij});\;\; \Phi_{ij}P_j = (\Phi_{xij}P_{xj};\Phi_{yij}P_{jy}) \tag{A.5.1}$$

$$\Omega_{ij} = (\Omega_{xij};\Omega_{yij});\;\; \Omega_{ij}D_j = (\Omega_{xij}D_{xj};\Omega_{yij}D_{jy}) \tag{A.5.2}$$

$$\Phi_{xij} = d_{xi}\delta_{ij}\;;\;\; \Phi_{yij} = d_{yi}\delta_{ij} \tag{A.5.3}$$

$$\Omega_{xij} = c_{xi}\delta_{ij}\;;\;\; \Omega_{xij} = c_{xi}\delta_{ij} \tag{A.5.4}$$

$$\Phi_{xij}P_{jx}(t,\boldsymbol{z}) = d_{xi}\delta_{ij}P_{xj}(t,\boldsymbol{z}) = d_{xi}P_{xi}(t,\boldsymbol{z})\;;\;\Phi_{xij}P_{jx}(t,\boldsymbol{z}) = d_{yi}P_{yi}(t,\boldsymbol{z}) \tag{A.5.5}$$

$$\Omega_{xij}D_{xj}(t,\boldsymbol{z}) = c_{xi}\delta_{ij}D_{xj}(t,\boldsymbol{z}) = c_{xi}D_{xi}(t,\boldsymbol{z})\;;\;\Omega_{yij}D_{yj}(t,\boldsymbol{z}) = c_{yi}D_{yi}(t,\boldsymbol{z}) \tag{A.5.6}$$

Thus equations (5.3; 5.4) take form *(i=1,..n)*:

$$\frac{\partial P_{xi}}{\partial t} + \nabla\cdot(\boldsymbol{v}\,P_{xi}) = c_{xi}D_{xi}(t,\boldsymbol{z})\;\;;\;\frac{\partial P_{yi}}{\partial t} + \nabla\cdot(\boldsymbol{v}\,P_{yi}) = c_{yi}D_{yi}(t,\boldsymbol{z}) \tag{A.6.1}$$

$$\frac{\partial D_{xi}}{\partial t} + \nabla\cdot(\boldsymbol{u}\,D_{xi}) = d_{xi}P_{xi}(t,\boldsymbol{z})\;\;;\;\frac{\partial D_{iy}}{\partial t} + \nabla\cdot(\boldsymbol{u}\,D_{yi}) = d_{yi}P_{yi}(t,\boldsymbol{z}) \tag{A.6.2}$$

To derive equations on aggregate impulses *P(t)* and *D(t)* (5.4.2; 5.4.3) and their components $P_{xi}$, $P_{yi}$, $D_{xi}$, $D_{yi}$ let's take integral by *dz=dxdy* of equation (A.5.3):

$$\frac{d}{dt}P_{xi}(t) = \frac{d}{dt}\int d\boldsymbol{z}\,P_{xi}(t,\boldsymbol{z}) = -\int d\boldsymbol{z}\,\nabla\cdot(\boldsymbol{v}\,P_{xi}) + c_{xi}\int d\boldsymbol{z}\,D_{xi}(t,\boldsymbol{z}) \tag{A.6.3}$$

Due to relations (3.4;3.5) and similar relations concern impulses $D_{xi}$, $D_{yi}$ obtain

$$P_{xi}(t) = \int d\boldsymbol{z}\,P_{xi}(t,\boldsymbol{z}) = C(t)v_{xi}(t);\; P_{yi}(t) = \int d\boldsymbol{z}\,P_{yi}(t,\boldsymbol{z}) = C(t)v_{yi}(t) \tag{A.6.3.1}$$

$$D_{xi}(t) = \int dz\, D_{xi}(t, z) = LR(t)u_{xi}(t); \quad D_{yi}(t) = \int dz\, D_{yi}(t, z) = LR(t)u_{yi}(t) \quad (A.6.3.2)$$

Due to same reasons as (A.2.1) first integral in the right side (A.6.3) equals zero and equations (A.6.1; A.6.2) takes form ($i=1,..n$):

$$\frac{d}{dt}P_{xi}(t) = c_{xi}D_{xi}(t) \; ; \; \frac{d}{dt}D_{xi}(t) = d_{xi}P_{xi}(t) \quad (A.6.4)$$

$$\frac{d}{dt}P_{yi}(t) = c_{yi}D_{yi}(t) \; ; \; \frac{d}{dt}D_{yi}(t) = d_{yi}P_{yi}(t) \quad (A.6.5)$$

Due to (A.1) impulses $P_{xi}(t)$, $P_{yi}(t)$, $D_{xi}(t)$, $D_{yi}(t)$ along each risk axes can't keep definite sign as in such a case they will reach max or min borders (A.1). Thus impulses along each axes must fluctuate and equations (A.6.4; A.6.5) describe simplest harmonique oscillations with frequencies $\omega_i$, $v_i$:

$$\omega_i^2 = -c_{xi}d_{xi} > 0 \; ; \; v_i^2 = -c_{yi}d_{yi} > 0 \; ; \; i = 1,..n \quad (A.6.6)$$

$$\left[\frac{d^2}{dt^2} + \omega_i^2\right]P_{xi}(t) = 0 \; ; \; \left[\frac{d^2}{dt^2} + \omega_i^2\right]D_{xi}(t) = 0 \quad (A.6.7)$$

$$\left[\frac{d^2}{dt^2} + v_i^2\right]P_{yi}(t) = 0 \; ; \; \left[\frac{d^2}{dt^2} + v_i^2\right]D_{yi}(t) = 0 \quad (A.6.8)$$

Equations (A.6.6-A.6.8) describe simple harmonique oscillations of impulses $P_{xi}(t)$, $P_{yi}(t)$, $D_{xi}(t)$, $D_{yi}(t)$ along each risk axes with different frequencies $\omega_i$, $v_i$ for $i=1,..n$. Frequencies $\omega_i$, $i=1,..n$ describe possible oscillations related to fluctuations of transactions from Creditors along coordinates $x=(x_1,..x_n)$. Frequencies $v_i$, $i=1,..n$ describe oscillations due to Borrowers along coordinates $y=(y_1,..y_n)$. Solutions of (A.6.7-8) have form:

$$P_{xi}(t) = P_{xi}(1)\sin\omega_i t + P_{xi}(2)\cos\omega_i t\,;\, P_{yi}(t) = P_{yi}(1)\sin v_i t + P_{yi}(2)\cos v_i t \quad (A.6.9)$$

$$D_{xi}(t) = D_{xi}(1)\sin\omega_i t + D_{xi}(2)\cos\omega_i t\,;\, D_{yi}(t) = D_{yi}(1)\sin v_i t + D_{yi}(2)\cos v_i t \quad (A.6.10)$$

Thus motions of Creditors and Borrowers on e-space induce oscillations (A.6.9-10) of macro transactions impulses with different frequencies $\omega_i$ and $v_i$ along risk axes $x_i$ or $y_i$. To derive equations on $Pz(t)$ and $Dz(t)$ determined by (A.3.1;A.3.2) let's define their components $Pz_{xi}(t);Pz_{yi}(t);Dz_{xi}(t);Dz_{yi}(t)$ as:

$$Pz_{xi}(t) = \int d\mathbf{x}d\mathbf{y}\, x_i P_{xi}(t, \mathbf{x}, \mathbf{y}) \; ; \; Pz_{yi}(t) = \int d\mathbf{x}d\mathbf{y}\, y_i P_{yi}(t, \mathbf{x}, \mathbf{y}) \quad (A.7.1)$$

$$Dz_{xi}(t) = \int d\mathbf{x}d\mathbf{y}\, x_i D_{xi}(t, \mathbf{x}, \mathbf{y}) \; ; \; Dz_{yi}(t) = \int d\mathbf{x}d\mathbf{y}\, y_i D_{yi}(t, \mathbf{x}, \mathbf{y}) \quad (A.7.2)$$

Relations (A.3.1;A.3.2) can be presented as:

$$Pz(t) = \sum_{i=1}^{n} Pz_{xi}(t) + \sum_{i=1}^{n} Pz_{yi}(t) \quad (A.7.3)$$

$$Dz(t) = \sum_{i=1}^{n} Dz_{xi}(t) + \sum_{i=1}^{n} Dz_{yi}(t) \quad (A.7.4)$$

To define equations on $Pz_{xi}(t)$, $Pz_{yi}(t)$, $Dz_{xi}(t)$, $Dz_{yi}(t)$ use equations (A.6.1 ; A.6.2). Let's multiply equations (A.6.1) by $x_i$ and take integral by $d\mathbf{x}d\mathbf{y}$

$$\frac{d}{dt}Pz_{xi}(t) = \frac{d}{dt}\int d\mathbf{x}d\mathbf{y}\, x_i P_{xi}(t, \mathbf{x}, \mathbf{y}) = -\int d\mathbf{x}d\mathbf{y}\, x_i \nabla\cdot(\mathbf{v}\, P_{xi}) + c_{xi}\int d\mathbf{x}d\mathbf{y}\, x_i D_{xi}(t, \mathbf{z})$$

$$\int dxdy\, x_i \nabla \cdot (\mathbf{v}\, P_{xi})$$

$$= \int dx_{k\neq i} d\mathbf{y} \int dx_i\, x_i \frac{\partial}{\partial x_i}(v_{xi} P_{xi}) + \int dx_i\, x_i \int dx_{k\neq i} d\mathbf{y} \frac{\partial}{\partial x_{k\neq i}}(v_{xk\neq i} P_{xi})$$

Second integral equals zero due to same reasons as (A.2.1). Let's take first integral by parts:

$$\int dx_i\, x_i \frac{\partial}{\partial x_i}(v_{xi} P_{xi}) = \int dx_i \frac{\partial}{\partial x_i}(x_i v_{xi} P_{xi}) - \int dx_i\, v_{xi} P_{xi}$$

First integral in the right side equals zero and we obtain:

$$\int dxdy\, x_i \nabla \cdot (\mathbf{v}\, P_{xi}) = -\int dxdy\, v_i P_{xi} = -\int dxdy\, v_i^2(t, \mathbf{x}, \mathbf{y}) CL(t, \mathbf{x}, \mathbf{y}) \qquad (A8.1)$$

Let's denote as

$$ECx_i(t) = \int dxdy\, v_{xi}^2(t, \mathbf{x}, \mathbf{y}) CL(t, \mathbf{x}, \mathbf{y});\ ECy_i(t) = \int dxdy\, v_{yi}^2(t, \mathbf{x}, \mathbf{y}) CL(t, \mathbf{x}, \mathbf{y}) \quad (A.8.2)$$

$$ERx_i(t) = \int dxdy\, u_{xi}^2(t, \mathbf{x}, \mathbf{y}) LR(t, \mathbf{x}, \mathbf{y});\ ERy_i(t) = \int dxdy\, u_{yi}^2(t, \mathbf{x}, \mathbf{y}) LR(t, \mathbf{x}, \mathbf{y}) \quad (A.8.3)$$

Thus equations on $Pz_{xi}(t)$, $Pz_{yi}(t)$, $Dz_{xi}(t)$, $Dz_{yi}(t)$ take form:

$$\frac{d}{dt} Pz_{xi}(t) = ECx_i(t) + c_{xi} Dz_{xi}(t)\ ;\ \frac{d}{dt} Dz_{xi}(t) = ERx_i(t) + d_{xi} Pz_{xi}(t)$$

$$\frac{d}{dt} Pz_{yi}(t) = ECy_i(t) + c_{yi} Dz_{yi}(t)\ ;\ \frac{d}{dt} Dz_{yi}(t) = ERy_i(t) + d_{yi} Pz_{yi}(t)$$

Due to relations (A.6.6) above equations on $Pz_{xi}(t)$, $Pz_{yi}(t)$, $Dz_{xi}(t)$, $Dz_{yi}(t)$ can be presented as:

$$\left[\frac{d^2}{dt^2} + \omega_i^2\right] Pz_{xi}(t) = \frac{d}{dt} ECx_i(t) + c_{xi} ERx_i(t) \qquad (A.8.4)$$

$$\left[\frac{d^2}{dt^2} + \omega_i^2\right] Dz_{xi}(t) = \frac{d}{dt} ERx_i(t) + d_{xi} ECx_i(t) \qquad (A.8.5)$$

$$\left[\frac{d^2}{dt^2} + v_i^2\right] Pz_{yi}(t) = \frac{d}{dt} ECy_i(t) + c_{yi} ERy_i(t) \qquad (A.8.6)$$

$$\left[\frac{d^2}{dt^2} + v_i^2\right] Dz_{yi}(t) = \frac{d}{dt} ERy_i(t) + d_{yi} ECy_i(t) \qquad (A.8.7)$$

To close system of ODE (A.4; A.8.4-7) let's derive equations on $ECx_i(t)$, $ECy_i(t)$, $ERx_i(t)$, $ERy_i(t)$. Let's outline that relations (A.8.2; A.8.3) are proportional to product of squares of velocities are *alike to* of energy of flow with velocity $v_{xi}$ or $v_{yi}$

$$EC(t) = \int dxdy\, v^2(t, \mathbf{x}, \mathbf{y}) CL(t, \mathbf{x}, \mathbf{y}) = C(t) v^2(t) = \sum_{i=1}^n ECx_i(t) + ECy_i(t) \quad (A.8.8)$$

$$ER(t) = \int dxdy\, u^2(t, \mathbf{x}, \mathbf{y}) LR(t, \mathbf{x}, \mathbf{y}) = LR(t) u^2(t) = \sum_{i=1}^n ERx_i(t) + ERy_i(t) \quad (A.8.9)$$

Let's regard $ECx_i(t)$ and $ECy_i(t)$ as components of $EC(t)$ along each axes $x_i$ and $y_i$. Relations (A.8.8 - 9) are *alike to* kinetic energy of particle with mass $C(t)$ and square velocity $v^2(t)$ and for convenience let's call $EC(t)$ and $ER(t)$ further as energies of corresponding flows. These similarities have no further analogies as no conservation laws on factors $EB(t)$ and $ER(t)$ exist. Equations on $ECx_i(t,z)$ and $ECy_i(t,z)$ take form similar to (4.1):

$$\frac{\partial}{\partial t} ECx_i(t, \mathbf{z}) + \nabla \cdot (\mathbf{v}\, ECx_i) = QECx_i\ ;\ \frac{\partial}{\partial t} ECy_i(t, \mathbf{z}) + \nabla \cdot (\mathbf{v}\, ECy_i) = QECy_i \qquad (A.9.1)$$

$$\frac{\partial}{\partial t} ERx_i(t,\mathbf{z}) + \nabla \cdot (\mathbf{u}\, ERx_i) = QERx_i \; ; \quad \frac{\partial}{\partial t} ERy_i(t,\mathbf{z}) + \nabla \cdot (\mathbf{u}\, ERy_i) = QERy_i \quad (A.9.2)$$

Let's propose that factors $QECx_i$ take form of diagonal matrix as:

$$QECx_i = \mathrm{M}_{xij} ERx_j = \mu_{xi}\, ERx_i \;; \quad \mathrm{M}_{xij} = \mu_{xi} \delta_{ij} \tag{A.9.3}$$

$$QECy_i = \mathrm{M}_{yij} ERy_j = \mu_{yi}\, ERy_i \;; \quad \mathrm{M}_{yij} = \mu_{yi} \delta_{ij} \mu_{yi} \tag{A.9.4}$$

$$QERx_i = \mathrm{N}_{xij} ECx_j = \eta_{xi} ECx_i \;; \quad \mathrm{N}_{xij} = \eta_{xi} \delta_{ij} \tag{F.9.5}$$

$$QERy_i = \mathrm{N}_{yij} ECy_j = \eta_{yi} ECy_i \;; \quad \mathrm{N}_{yij} = \eta_{yi} \delta_{ij} \tag{A.9.6}$$

$$\gamma_{xi}^2 = \mu_{xi} \eta_{xi} > 0 \;; \quad \gamma_{yi}^2 = \mu_{yi} \eta_{yi} > 0 \tag{A.9.7}$$

Similar to derivation of equations on impulses $P_{xi}(t)$, $P_{yi}(t)$, $D_{xi}(t)$, $D_{yi}(t)$ (A.6.4-A.6.8) equations (A.9.1-7) give equations on $ECx_i(t)$, $ECy_i(t)$, $ERx_i(t)$, $ERy_i(t)$:

$$\left[\frac{d^2}{dt^2} - \gamma_{xi}^2\right] EC_{xi}(t) = 0 \;; \quad \left[\frac{d^2}{dt^2} - \gamma_{xi}^2\right] ER_{xi}(t) = 0 \tag{A.10.1}$$

$$\left[\frac{d^2}{dt^2} - \gamma_{yi}^2\right] EC_{yi}(t) = 0 \;; \quad \left[\frac{d^2}{dt^2} - \gamma_{yi}^2\right] ER_{yi}(t) = 0 \tag{A.10.2}$$

Economic meaning of (A.9.1-A.9.7) is as follows: "energies" $ECx_i(t)$, $ECy_i(t)$, $ERx_i(t)$, $ERy_i(t)$ grow up or decay in time by exponent $exp(\gamma_{xi} t)$ and $exp(\gamma_{yi} t)$ that can be different for each risk axis $i=1,..n$. Here $\gamma_{xi}$ define exponential growth or decay in time of $ECx_i(t)$ induced by motion of Creditors along axes $x_i$ and $\gamma_{yi}$ same time describe exponential growth or decrease in time of $ECy_i(t)$ induced by motion of Borrowers along axes $y_i$. The same valid for $ERx_i(t)$, $ERy_i(t)$ respectively. Equations (A.4; A.8.4-7; A.10.1-2) describe a closed system of ODE that models time evolution of aggregate variables $C(t)$, $LR(t)$, $P_{zxi}(t)$, $P_{zyi}(t)$, $D_{zxi}(t)$, $D_{zyi}(t)$, $ECx_i(t)$, $ECy_i(t)$, $ERx_i(t)$, $ERy_i(t)$ and solutions (A.4; A.8.4-7; A.10.1-2) have form:

$$EC_{xi}(t) = EC_{xi}(1) \exp \gamma_{xi} t + EC_{xi}(2) \exp -\gamma_{xi} t$$

$$EC_{yi}(t) = EC_{yi}(1) \exp \gamma_{yi} t + EC_{yi}(2) \exp -\gamma_{yi} t$$

$$ER_{xi}(t) = ER_{xi}(1) \exp \gamma_{xi} t + ER_{xi}(2) \exp -\gamma_{xi} t$$

$$ER_{yi}(t) = ER_{yi}(1) \exp \gamma_{yi} t + ER_{yi}(2) \exp -\gamma_{yi} t$$

$$Pz_{xi}(t) = Pz_{xi}(1) \sin \omega_i t + Pz_{xi}(2) \cos \omega_i t + Pz_{xi}(3) \exp \gamma_{xi} t + Pz_{xi}(4) \exp -\gamma_{xi} t$$

$$Pz_{yi}(t) = Pz_{yi}(1) \sin v_i t + Pz_{yi}(2) \cos v_i t + Pz_{yi}(3) \exp \gamma_{yi} t + Pz_{yi}(4) \exp -\gamma_{yi} t$$

$$Dz_{xi}(t) = Dz_{xi}(1) \sin \omega_i t + Dz_{xi}(2) \cos \omega_i t + Dz_{xi}(3) \exp \gamma_{xi} t + Dz_{xi}(4) \exp -\gamma_{xi} t$$

$$Dz_{yi}(t) = Dz_{yi}(1) \sin v_i t + Dz_{yi}(2) \cos v_i t + Dz_{yi}(3) \exp \gamma_{yi} t + Dz_{yi}(4) \exp -\gamma_{yi} t$$

Total Credits $C(t)$ as solution of (A.4; A.7.4) have form:

$$C(t) = C(0) + a \sum_{i=1}^{n} [\, C_{xi}(1) \sin \omega_i t + C_{xi}(2) \cos \omega_i t + C_{yi}(3) \sin v_i t + C_{yi}(4) \cos v_i t\,] +$$
$$a \sum_{i=1}^{n} [C_{xi}(5) \exp \gamma_{xi} t + C_{xi}(6) \exp -\gamma_{xi} t + C_{yi}(7) \exp \gamma_{yi} t + C_{yi}(8) \exp -\gamma_{yi} t] \quad (A.11)$$

Simple but long relations define constants $C_{xi}(j)$, $C_{yi}(j)$, $j=0,..8$ that are determined by initial values and equations (A.4; A.8.4-7; A.10.1-2) and we omit them here. Similar relations are valid for total rate of Loan-Repayment *LR(t)* (5.4.1). Solutions (A.10) allow obtain simple relations on macro Credits *MC(t)* (3.10; 3.11).